\documentclass[12pt,preprint]{aastex}

\newcommand{\hi}{H\,I}

\shorttitle{M31-M33 \hi\ Bridge}
\shortauthors{Lockman et al.}

\begin{document}

\title{The Neutral Hydrogen Bridge between M31 and M33}

\author{Felix J. Lockman}
\affil{National Radio Astronomy Observatory \altaffilmark{1}, Green Bank, WV 24944}
\email{jlockman@nrao.edu}

\author{Nicole L. Free and Joseph C. Shields}
\affil{Dept. of Physics and Astronomy, Ohio University, Athens OH 45701}

\altaffiltext{1}{The National Radio Astronomy Observatory is a facility of the National Science Foundation operated under a cooperative agreement by Associated Universities, Inc.}

\begin{abstract}

  The Green Bank Telescope has been used to search for 21cm \hi\
emission over a large area between the galaxies M31 and M33 in an
attempt to confirm at $9\farcm1$ angular resolution the detection by
\citet{BraunThilker2004} of a very extensive neutral gas ``bridge''
between the two systems at the level N$_{\rm HI} \approx 10^{17}$
cm$^{-2}$.  We detect \hi\ emission at several locations up to 120 kpc
in projected distance from M31, at least half the distance to M33,
with velocities similar to that of the galaxies, confirming the
essence of the Braun \& Thilker discovery.  The \hi\ does not appear
to be associated with the extraplanar high-velocity clouds of either
galaxy.  In two places we measure N$_{\rm HI} > 3 \times 10^{18}$
cm$^{-2}$, indicative of concentrations of \hi\ with $\sim 10^5$
M$_{\odot}$ on scales $\lesssim 2$ kpc, but over most of the field we
have only $5\sigma$ upper limits of N$_{\rm HI} \leq 1.4 \times
10^{18}$ cm$^{-2}$.  In very deep measurements in two directions \hi\
lines were detected at a few $10^{17}$ cm$^{-2}$.  The absence of
emission at another location to a $5\sigma$ limit N$_{\rm HI} \leq 1.5
\times 10^{17}$ cm$^{-2}$ suggests that the \hi\ bridge is either
patchy or confined to within $\sim 125$ kpc of M31.  The measurements
also cover two of M31's dwarf galaxies, And~II and And~XV, but in
neither case is there evidence for associated \hi\ at the $5\sigma$
level of $1.4 \times 10^4$ M$_{\odot}$ for And~II, and $9.3 \times
10^3$ M$_{\odot}$ for And~XV.

\end{abstract}



\section{Introduction}

The larger spiral galaxies of the Local Group, the Milky Way, M31 and M33, are growing through mergers with smaller systems.  M31 and M33 contain numerous stellar streams \citep{Ibata2001, Ferguson2002, McConnachie2009}, and the Milky Way is currently accreting the Sagittarius dwarf galaxy,  
the Smith cloud, and probably the Magellanic Stream as well  \citep{ Matthewson1974,Ibata1994, Putman2003,Lockman2008,McClure-Griffiths2008, Stanimirovic2008, Nidever2010}.   M31, M33, and the Milky Way are also surrounded by clouds of neutral and ionized gas -- the ``high velocity clouds'' (HVCs) -- that are not identifiable with any stellar system and may be dark matter subhalos,  material stripped from satellites, accretion from a hot halo or the intergalactic medium, or something else entirely  \citep{wvw97,Thilker2004, Westmeier2008, Grossi2008, Shull2009, Nichols2009}.   

Braun \& Thilker (2004, hereafter B\&T) mapped the \hi\ over a large area around M31 and M33, and reported the detection of extremely faint 21cm \hi\ emission at the level log(N$_{\rm HI}) \approx 17.0$ cm$^{-2}$  that formed a partial bridge about 200 kpc in extent between the two galaxies. This gas lies well outside  each galaxy's HVC system and has been interpreted as the neutral component of an intergalactic filament, or the remnant of a past encounter between M31 and M33 \citep[B\&T;][]{Bekki2008}.   Discussion of the likelihood and consequences of  an interaction between  M33 and M31 are given in recent papers by \citet{Putman2009},  \citet{Davidge2011}, and \citet{Peebles2011}. For convenience we will refer to the structure reported by B\&T as the M31 ``bridge''.  

 B\&T made their measurements using the Westerbork Synthesis Radio Telescope configured as a group of single dishes to obtain very high sensitivity to low surface-brightness \hi\ emission, but at the expense of angular resolution.    They further smoothed the data in angle and velocity to  detect  this extremely weak emission more easily: their final maps had 49\arcmin\ angular resolution, equivalent to 11 kpc at the distance of M31, and a velocity resolution of 17 km s$^{-1}$. Even so, much of the signal they reported was just  $2\sigma$ to $3\sigma$  above the noise.  Although they obtained a spectrum with the Green Bank Telescope (GBT) that was consistent with their Westerbork ``single dish'' measurements at one location, over much of the field the detections are only marginally significant.

Subsequently, \citet{Putman2009} questioned the existence of
this \hi\ bridge, noting its absence from the immediate vicinity of M33
at the level log(N$_{\rm HI})\gtrsim 18.0$ in data from the Arecibo radio telescope, and suggested that the B\&T result might be blended Galactic \hi\ mistakenly attributed to M31 or M33.  

\hi\ is typically the most massive component of tidal tails \citep[e.g.,][]{Duc2011}, so  the existence of a neutral hydrogen bridge between M31 and M33 may be key in understanding the evolution of both galaxies.  To determine the reality of the bridge,  and to study its  properties at significantly higher angular resolution than B\&T,  we have observed a large area between M33 and M31 
using the Green Bank Telescope  at a factor $\sim5$ better resolution in angle and velocity. As importantly, the very clean optics of the GBT \citep{Boothroyd2011} greatly  reduces the chances that any  faint 21cm detection is a spurious signal  that originates from the  bright \hi\  disks of  M31 or M33 and has entered the receiver through a sidelobe.

Spectra in the map presented here have  a $5\sigma$
sensitivity limit of log(N$_{\rm HI})\approx18.0$,  so  the gas in the B\&T bridge would not be detected if it is smoothly distributed on scales $\approx 1\degr\ (14$ kpc), but  our hope was that the bridge would have much brighter small-scale structure, unresolved by the $49\arcmin$ observations of 
B\&T, but well resolved by the $9\farcm1$ beam of the GBT.  In addition to the map, three directions were selected for very deep observations capable of detecting \hi\ emission at the level log(N$_{\rm HI})\approx 17.0$.  If the \hi\ structure between M31 and M33 is actually diffuse, i.e., not highly spatially structured, and at the brightness implied by the B\&T measurements, the emission would not be seen in the map, only in the deep pointings.

The goals of this project are thus: 1) to search a wide area between M31 and M33 for evidence of the \hi\ bridge arising from clumps in the gas, and  2) to measure the \hi\ at several locations at a sensitivity similar to that of B\&T but with much greater angular resolution as a check on the basic existence of the bridge.  Additional measurements were made toward two of M31's dwarf galaxies that lie within the mapped region.  The observations are discussed in \S\ref{sec:obs}, results toward the dwarfs And~II and And~XV are presented in \S\ref{sec:dwarfs},  \hi\ emission that is widespread across the field but probably not related to the M31 bridge is discussed in \S\ref{sec:v-200}, results that pertain to the \hi\ bridge are presented in \S\ref{sec:detections}, the present measurements are discussed in relation to the HVC systems of M31 and M33 in \S\ref{sec:HVCs} , and the concluding discussion is in    \S\ref{sec:discussion}.

\section{Observations}

 \label{sec:obs}

An area of approximately 50 square degrees between M31 and M33 was
observed with the 100-meter diameter Robert C. Byrd Green Bank
Telescope \citep{Prestage2009}, at an angular resolution of $9\farcm1$
using the L-band receiver, which has a typical system temperature at
zenith of 18 K in both linear polarizations. All observations were
made over a bandwidth of 12.5 MHz at a channel spacing of 0.16
km~s$^{-1}$.  During the data reduction spectra were smoothed to
coarser velocity resolution using a boxcar function and resampled.
Spectra were edited, then calibrated and corrected for stray radiation
as described by \citet{Boothroyd2011}, though at the velocities of
interest here stray radiation is negligible.  All intensities quoted
here are brightness temperatures (T$_{\rm b}$) averaged over the
$9\farcm1$ main beam, corrected for atmospheric absorption.  Three
data sets were acquired: maps, follow-up observations, and deep
pointings.  The $5\sigma$ detection limits for a 25 km~s$^{-1}$ (FWHM)
line are given in Table \ref{tab:table_sensitivity} for both \hi\
column density and \hi\ mass within the GBT beam.

The GBT is an extremely sensitive instrument for 21cm \hi\ spectroscopy. Early suggestions of $\approx 10\%$ ``gain fluctuations'' in 21cm spectra \citep{Robishaw2009} have not been confirmed in a series of rigorous checks \citep{Boothroyd2011}, while hundreds of thousands of \hi\ spectra have been measured by different groups with accuracies limited only by noise, or quantifiable effects such as instrumental baseline variations \citep[e.g.,][]{Lockman2005, Hogg2007, Nidever2010,  Chynoweth2011}.  Determinations of foreground Galactic  N$_{\rm HI}$ toward AGN using the GBT in the 21cm line are in excellent agreement  with values derived from Lyman-$\alpha$ absorption lines in the same direction \citep{Wakker2011}

The observations were made by Doppler correcting to a constant V$_{\rm LSR}$ at the center of each spectrum, and the final data cube was gridded in constant  V$_{\rm LSR}$ channels.  Conversion of velocities to Heliocentric and  Local Group Standard of Rest (LGSR) were made using the apex velocity and coordinates given by \citet{Karachentsev1996}.  For our data the uncertainties in the resulting V$_{\rm LGSR}$ are dominated by systematic uncertainties in the direction and velocity of the apex.

\subsection{Maps of the Area Between M31 and M33}
\label{sec:mapobs}

The GBT was used to map 
 21cm \hi\ emission  over a large area between M33 and M31 guided by the B\&T results.  For convenience, the survey region was divided in blocks of $2\degr \times 2\degr$.  Spacing between samples on the sky was $3\farcm5$, somewhat finer than the Nyquist sampling interval of $3\farcm6$ for the GBT's $9\farcm1$ beam.    Each position in a block was observed for 6 seconds, and each block was observed as many as six times.  Data were acquired during more than two dozen independent observing sessions over a period of 20 months.

	In-band frequency switching gave a useable velocity coverage between -600 and +470 km s$^{-1}$ (LSR).   The spectra were smoothed from their intrinsic velocity resolution of 0.16 km~s$^{-1}$ to 2.90 km~s$^{-1}$, and a 3rd order polynomial was fit to emission-free portions of each spectrum.  The data were assembled into a cube as described by Mangum et al. (2007) and \citet{Boothroyd2011}, and a third order polynomial was removed from each pixel. The spectra suffered from occasional narrow-band interference generated within the GBT receiver room.  These signals were stable in frequency and  appeared in only one spectral channel, so spectra were interpolated over the affected channels.   

Occasionally, during some of the frequency-switched scans, one of the linearly polarized receiver channels had a poor spectral baseline likely caused by out-of-band interference.  While it was possible in the ``follow-up'' and ``deep''  pointings to examine the data minute-by-minute for  problems, the data over the mapped region consists of many tens of thousands of short measurements,   and scrutiny of individual raw spectra was impractical.  For this reason we retained only data from the good receiver channel for use in the maps.  This problem did not appear in data taken by position switching.

\begin{figure}
\resizebox{1.0\columnwidth}{!}{%
\includegraphics{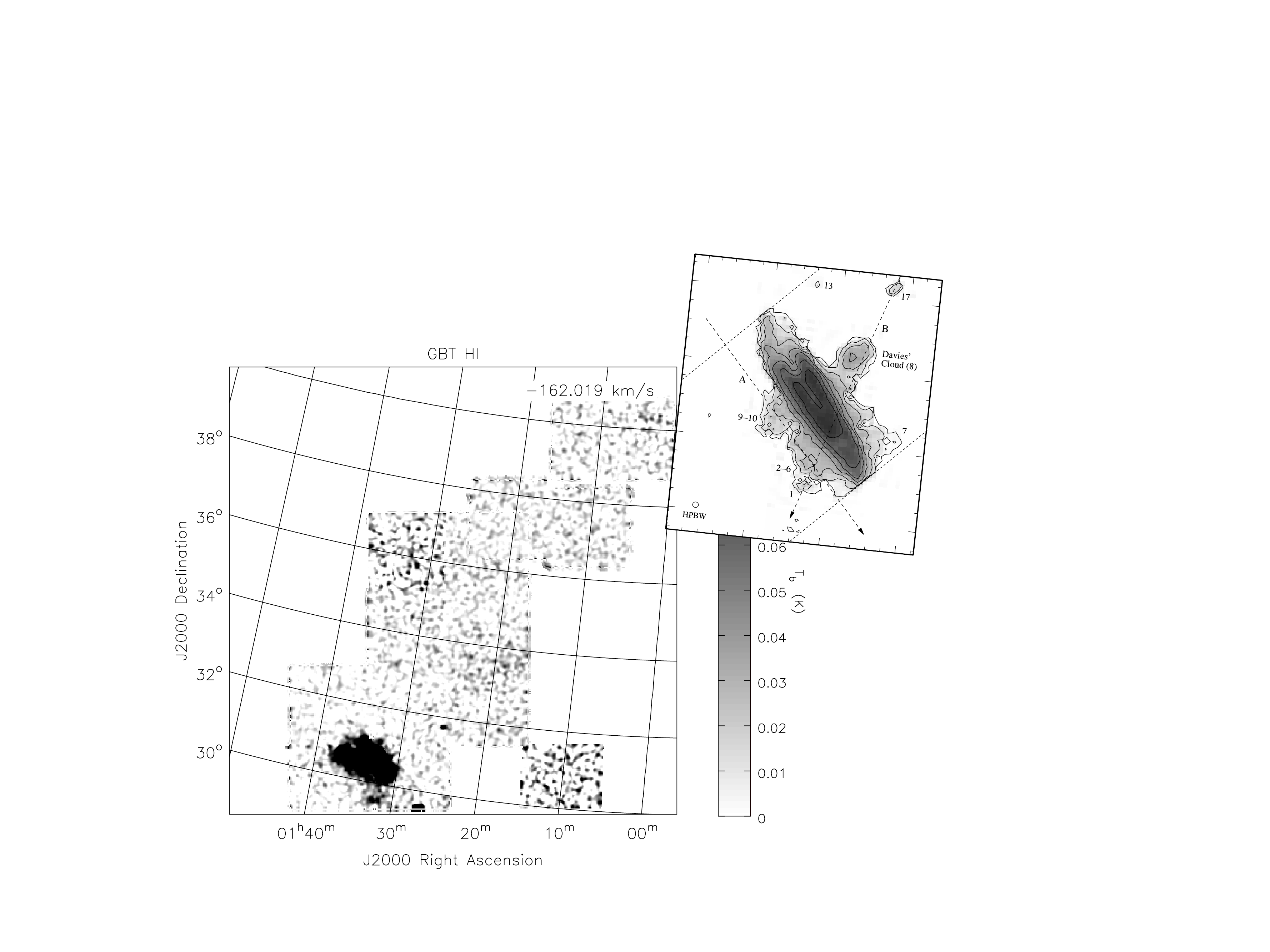} }
  \caption
  {The area covered by the present GBT observations at $9\farcm1$ resolution is shown in a channel map at an emission-free velocity to illustrate the noise, with an \hi\   image of M31 and its extended system of HVCs \citep[from][]{Westmeier2008} at the upper right in its relative location.  Our map overlaps some of the area covered by \citet{Westmeier2008} and includes a large part of the \hi\ bridge discovered by B\&T, but avoids M31's HVCs. The angular resolution of the GBT data is essentially identical to that shown in the M31 insert.   \hi\ emission from the galaxy  M33 appears at the lower left. 
}  
\label{fig:coverage}
\end{figure}

A channel map at a V$_{\rm LSR}$ that is free from 21cm  emission (except for M33) is shown in Figure
\ref{fig:coverage}.  The rms noise over most of the map in a 2.9 km~s$^{-1}$ channel is $\approx20$ mK except for the blocks at  J2000 $01^h33^m, +36\arcdeg$ and $01^h10^m, +31\arcdeg$, where because of reduced integration time the noise is  44 mK.  The area of this latter block, in the lower right of the figure, covers part of the Wright High-Velocity Cloud \citep{Wright1979} which confuses emission at negative velocities.  It was not included in this analysis.  For most of the map the $5\sigma$ detection limit for a 25 km s$^{-1}$ line (FWHM) is $1.5 \times 10^{18}$ cm$^{-2}$.  The characteristic angular size of the noise speckels  in the map gives an indication of the angular resolution, 9\farcm1.

 These mapping observations are 1.5-2.5  times more sensitive than the M31 HVC survey of \citet{Westmeier2008} which was made with a similar angular resolution over an area that overlaps the north-west portion of our map.  The Arecibo GALFA-HI observations of the M33 region \citep{Putman2009} have a much higher angular resolution than those presented here (3\farcm4 against 9\farcm1), but the GBT map spectra have a factor $\approx6$ lower noise, giving our observations essentially equal sensitivity for structures with an angular size $\lesssim3\arcmin$, and a factor of 6 more sensitivity to emission on scales $\gtrsim10\arcmin$.   The measurements of the galaxy M33 that resulted from our survey will be described elsewhere.

\subsection{Follow-up Pointed Observations}
\label{sec:follow-up}
Each location in the GBT map where 21cm emission was detected at V$_{LSR} \lesssim -200$ km s$^{-1}$  at the $\gtrsim3\sigma$ level was re-observed in pointed, frequency-switched observations for 8 minutes. These spectra have a $5\sigma$ sensitivity to a 25 km s$^{-1}$ line of $5 \times 10^{17}$  cm$^{-2}$.   
 Two of M31's dwarf galaxies that are within the boundaries of the map were also observed in this mode, And~II and And~XV.  Because And~II has a velocity similar to that of extended \hi\ features in its direction,   an area around that dwarf galaxy was mapped with one minute integrations.  This is discussed in \S~\ref{sec:AndII}.

\subsection{Deep Pointed Observations}
In addition to the map of the large area,  
three positions where the B\&T data showed detectable emission were selected for deep observations.
Two were chosen to be near localized peaks in the B\&T map where N$_{\rm HI} \approx 10^{17.5}$ cm$^{-2}$, the first at $01^h20^m, +37\arcdeg22\arcmin$, the second at $01^h00^m, +39\arcdeg30\arcmin$.  The third deep position was chosen to be near the southernmost tip of the B\&T bridge at $01^h20^m, +36\arcdeg00\arcmin$ where the B\&T map shows very weak \hi\  emission at the level of $10^{17}$ cm$^{-2}$.  

The $01^h20^m, +37\arcdeg22\arcmin$ position was observed in the usual frequency-switched mode for several minutes at the beginning of every observing session.  This gave a  cumulative integration time of more than three hours over the course of this project.   After normal processing, the emission-free regions were  fit with a 4th order polynomial and smoothed to a velocity resolution of 3.2 km~s$^{-1}$.  

The second set of deep pointed observations, toward $01^h00^m, +39\arcdeg30\arcmin$,  was made over six hours in a single evening in position-switched mode, using a reference position $3\degr$ to higher right ascension.   The data after calibration and smoothing in velocity were fit with a linear baseline.

The third position, at  $01^h20^m, +36\arcdeg00\arcmin$, was observed in a single 3 hour session, using in-band frequency switching with a setup similar to that of the follow-up observations.  These data were reduced in the standard way and fit with a 4th degree polynomial baseline, then smoothed to 3.2 km~s$^{-1}$ channel spacing.  

\begin{deluxetable}{lll}
  \tablecaption{$5\sigma$ Detection Limits\tablenotemark{1}
\label{tab:table_sensitivity}}
  \tablehead{ 
    \colhead{Data Set} & \colhead{N$_{\rm HI}$} & 
\colhead{M$_{\rm HI}$\tablenotemark{2}}  \\
  \colhead{ }  & \colhead{(cm$^{-2}$)} & \colhead{(M$_{\odot}$)} }
  \startdata
  Map & $1.5\times 10^{18}$ & $4.2 \times 10^4 $ \\
  Follow Up & $5.0\times 10^{17}$ & $ 1.4 \times 10^4$ \\
  Deep Pointings & $1.0-1.4\times 10^{17}$ & $2.8-4.1 \times 10^3 $\\
\enddata
\tablenotetext{1}{For a line width of 25 km s$^{-1}$ (FWHM).}
\tablenotetext{2}{Mass of \hi\ within a single GBT beam at 0.8 Mpc distance.}
\end{deluxetable}


\section{\hi\ in the M31 Dwarf Galaxies And~II and And~XV}
\label{sec:dwarfs}
Two dwarf spherodial galaxies associated with M31 lie within the boundaries of the survey, And~II and And~XV.   As it is of interest to see if there is any connection between M31's dwarf galaxies and the \hi\ bridge, we made pointed observations with the GBT toward both systems, and additionally, mapped a region around And~II.   No \hi\ was detected that could be associated unambiguously with either galaxy. The limits are given in Table \ref{tab:Satellites}.

\begin{deluxetable}{cccccclc}
\tabletypesize{\small}
\tablecaption{Observations of M31 Satellites
 \label{tab:Satellites}}
\tablehead{
    \colhead{J2000}  &  \colhead{V$_{\rm LSR}$} & \colhead{V$_{\rm LGSR}$} &\colhead{$\sigma_{\rm b}$\tablenotemark{1}} & \colhead{N$_{\rm HI}$ \tablenotemark{2}} & \colhead{M$_{\rm HI}$} & 
 \colhead{Object}  & \colhead{Ref}\\
\colhead{(hh:mm:ss.s dd:mm)}  & 
\colhead{(km s$^{-1}$)} & \colhead{(km s$^{-1}$)} & \colhead{(mK)} & \colhead{(10$^{17}$ cm$^{-2}$)} & 
\colhead{($10^3$ M$_{\odot}$)} 
\\}
  \startdata
01:14:18.7  +38:07        & $-322$ (1.4) &-79 (13)  & 4.4  & $<3.6$   & $<9.3$ & And~XV  & 1\\ 
01:16:29.8  +33:25        &  $-187$ (3.0) & +46 (14) & 9.4  & $<7.7$   &  $<14$   &  And~II &  2 \\
  \enddata\tablerefs{
(1) \citet{Tollerud2011}; (2) \citet{Cote1999}.}
\tablenotetext{1}{Noise in a 3.2 km~s$^{-1}$ channel}
\tablenotetext{2}{$5\sigma$ limits for a 25 km~s$^{-1}$ line width.}
\tablecomments{Upper limits are $5\sigma$. Quantities for And~II are derived from the difference between spectra toward that galaxy and an average of spectra at positions 10\arcmin\  to 15\arcmin\  away.}
\end{deluxetable}

\subsection{And~XV}
\label{sec:AndXV}
And~XV lies at a distance of 0.77 Mpc with a heliocentric radial velocity V$_{\rm HEL} =-323 \pm 1.4$ km~s$^{-1}$ (V$_{\rm LSR} =-322$ km~s$^{-1}$) and a half-light radius of less than 2\arcmin\ \citep{Letarte2009, Tollerud2011}.  
   No 21cm emission was detected at the appropriate velocity to a  $5\sigma$ upper limit N$_{\rm HI} \leq 3.6 \times 10^{17}$ cm$^{-2}$ for a 25 km~s$^{-1}$ line width.  At the distance of And~XV this is equivalent to a $5\sigma$ limit on the \hi\ mass of $\leq9.3 \times 10^3$ M$_{\odot}$ within a GBT beam.

\subsection{And~II}
\label{sec:AndII}
\begin{figure}
 \scalebox{0.5}[0.5]{\includegraphics{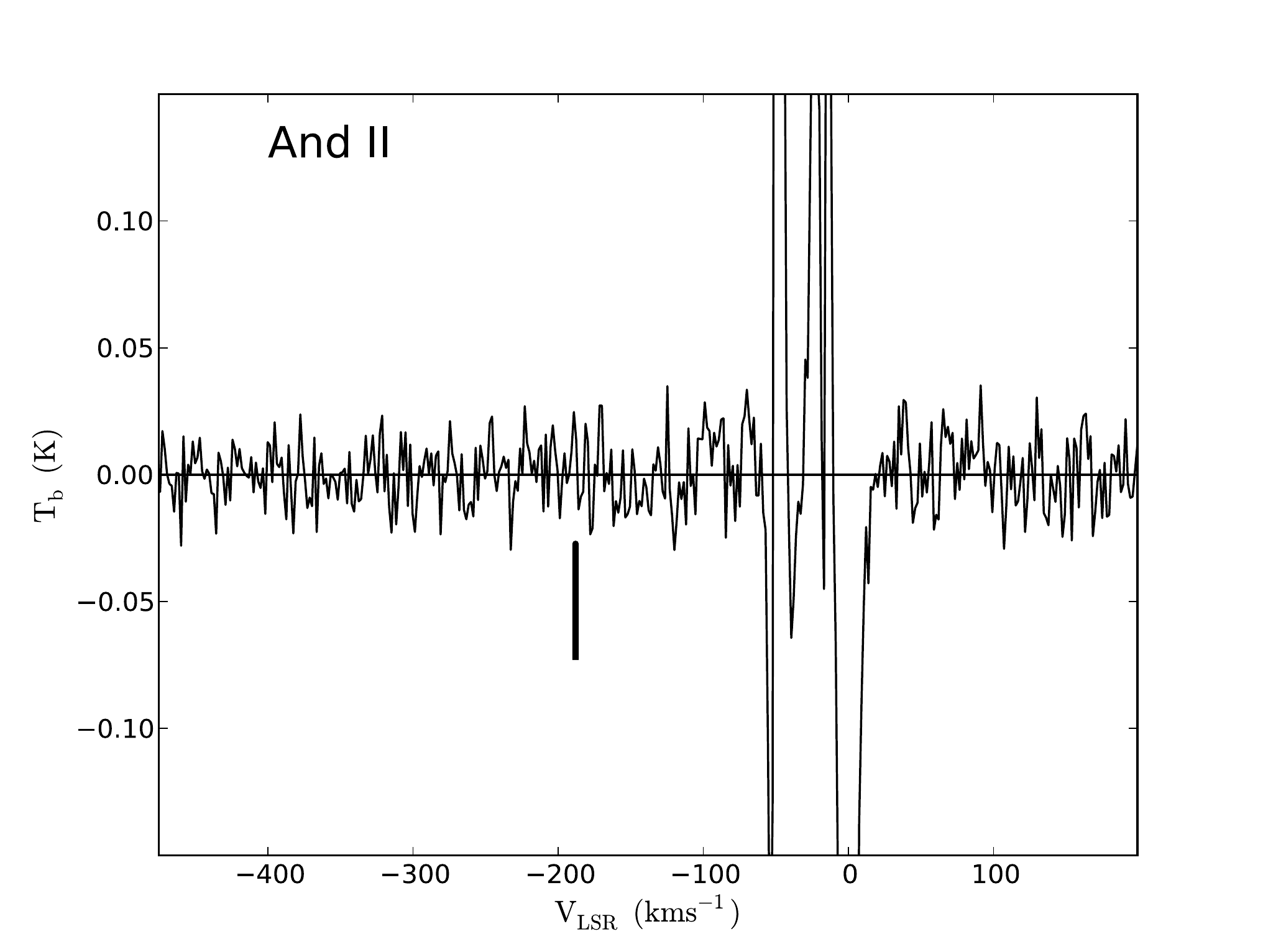}}
  \caption{
The difference between GBT 21cm \hi\ spectra toward the M31 dwarf galaxy And~II and the average of spectra $10\arcmin$ - $15\arcmin$ radially around the galaxy.  There is no significant emission at the velocity of the dwarf galaxy, V$_{\rm LSR} = -188$ km~s$^{-1}$ (marked with the bar), to a $5\sigma$ limit M$_{\rm HI} < 1.4 \times 10^4$ M$_{\odot}$ for a 25 km~s$^{-1}$ line.
}  
\label{fig:AndII_diff} 
\end{figure}

And~II, at a distance of 0.65 Mpc and V$_{\rm HEL} = -188$ km~s$^{-1}$
(V$_{\rm LSR} = -187$ km~s$^{-1}$) \citep{McConnachie2005,Cote1999},
is much closer to Galactic velocities than And~XV.  With an angular
size $\approx 4\arcmin$, the And II stars, like those of And XV, are
well covered by a GBT beam.  Spectra taken directly toward this dwarf
show weak \hi\ emission ($\approx 30$ mK) at the velocity of the
galaxy.  However, similar emission is seen at surrounding positions up
to several degrees away.  Figure~\ref{fig:AndII} shows the measured
\hi\ spectrum directly toward And~II and the average of several \hi\
spectra at positions offset in a ring 10\arcmin\ to 15\arcmin\ from
that galaxy.  The spectra have similar intensities at the velocity of
And II, regardless of whether the dwarf is in the beam or not.  The
difference between the \hi\ spectrum toward And~II and that toward the
reference positions is shown in Fig.~\ref{fig:AndII_diff}.  Any \hi\
coincident in position and velocity with the stellar component of this
dwarf must have a $5\sigma$ limit on N$_{\rm HI} \leq 7.7 \times
10^{17}$ cm$^{-2}$ for a 25 km~s$^{-1}$ line, with an associated
$5\sigma$ \hi\ mass limit M$_{\rm HI} \leq 1.4 \times 10^4$
M$_{\odot}$.


\section{\hi\ at $ -200 \leq V_{LSR} \leq -100$ km~s$^{-1}$}
\label{sec:v-200}

\begin{figure}
 \scalebox{0.5}[0.5]{\includegraphics{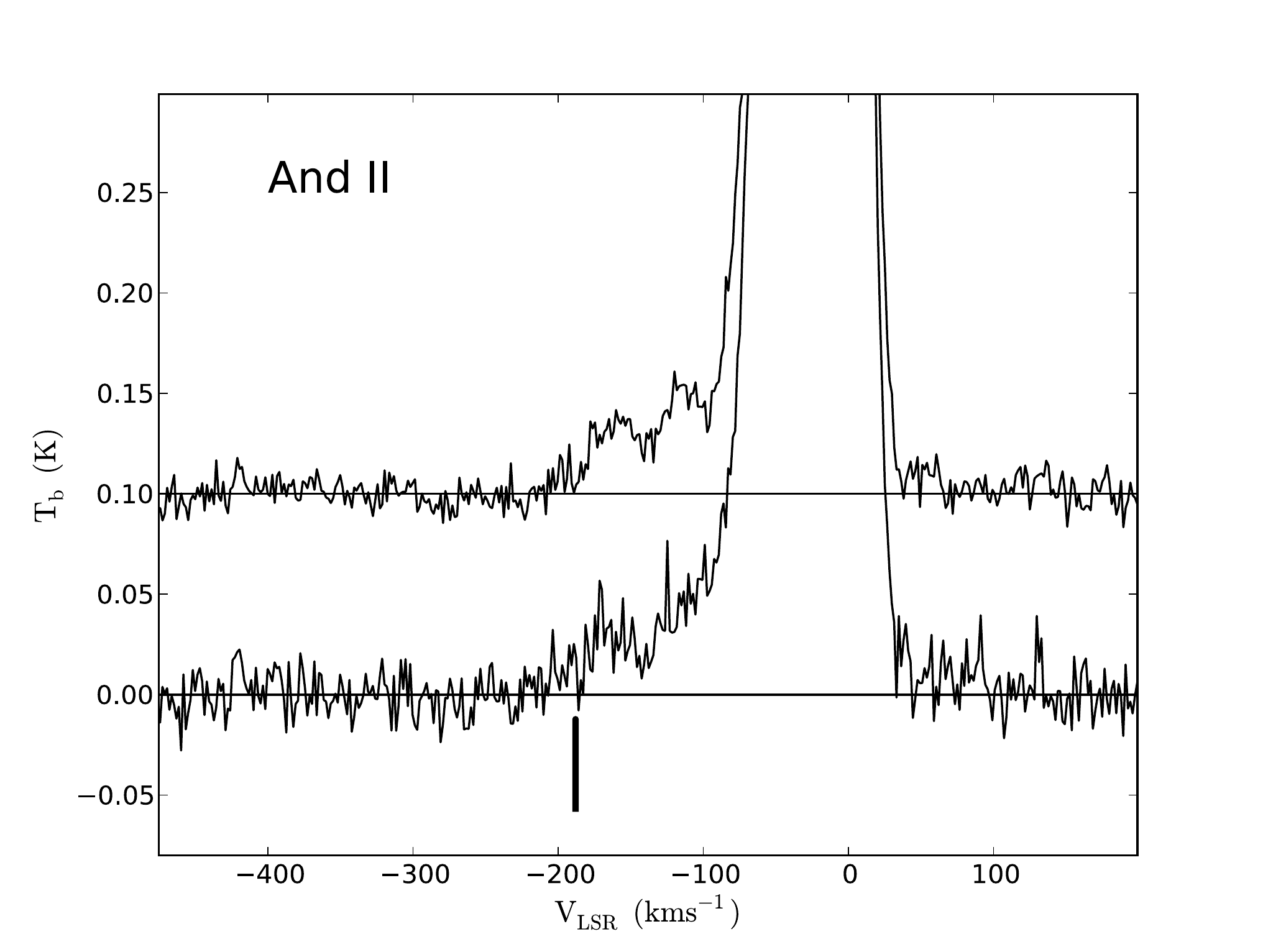}}
  \caption{
GBT 21cm spectrum of \hi\ directly toward And~II (lower curve) and averaged over a number of positions 10\arcmin\  to 15\arcmin\ away from it (upper curve).  The dwarf galaxy has a   V$_{\rm LSR} = -188$ km~s$^{-1}$ (marked with the vertical bar), a velocity where there  is \hi\ emission over a considerable area of the sky. 
}  
\label{fig:AndII} 
\end{figure}

\begin{figure}
 \scalebox{0.65}[0.65]{\includegraphics{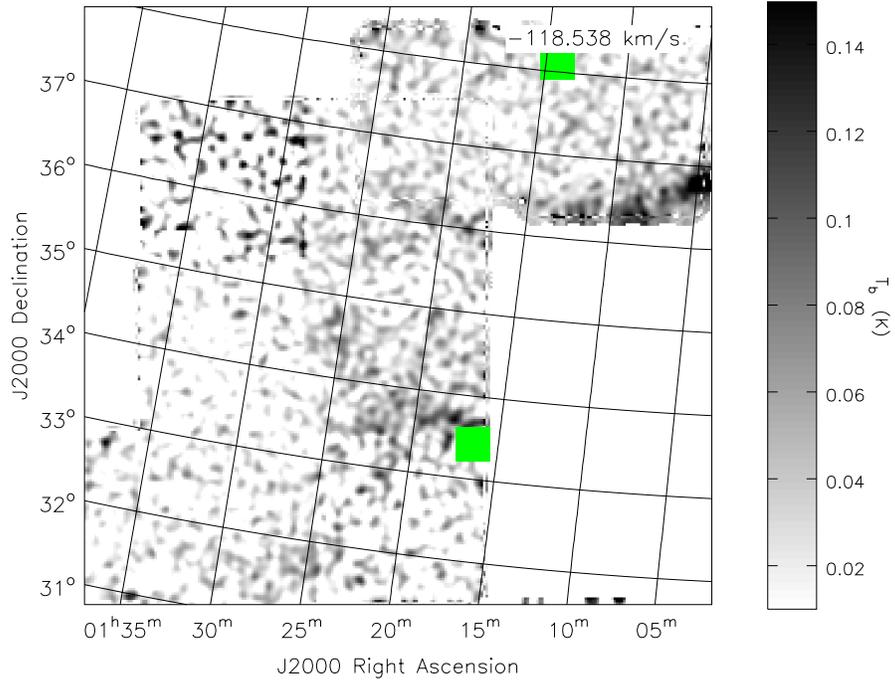}}
  \caption{
The western region of our map at V$_{\rm LSR}= -118.5$ km~s$^{-1}$ showing the extent of the emission near -120 km~s$^{-1}$.  Filled rectangles show the location of the dwarf galaxies And~XV to the north and And~II to the south.
}  
\label{fig:map-120_GBT} 
\end{figure}

Over the western portion of the mapped region there is \hi\ at $-150 \leq {\rm V_{ LSR}} \leq -100$ km~s$^{-1}$ in addition to the gas near And~II that has $-200 \leq $ V$_{\rm LSR} \leq -150$ km~s$^{-1}$ (both components are shown in the region of And~II in Fig.~\ref{fig:AndII}). The distribution of the -120 km~s$^{-1}$ component over the mapped region is shown in Fig.~\ref{fig:map-120_GBT}.  Much lower resolution, incompletely sampled data from the LAB survey \citep{Kalberla2005} show that this material lies in a long stream approximately along the great circle that connects M31 and M33.  Its velocity with respect to the LGSR is V$_{\rm LGSR} > +100$ km~s$^{-1}$, quite discrepant from that of M31, M33, and the \hi\ bridge, which over the region of this emission has V$_{\rm LGSR} \lesssim 0$ km~s$^{-1}$ (\S\ref{sec:detections}).  Although in the existing data there are suggestions of a spatial correlation of this gas with the galaxies, its discrepant velocity indicates that it must have an origin quite different from that of the bridge. 


\section{Detections of the M31-M33 HI Bridge}
\label{sec:detections}

\begin{figure}
 \scalebox{0.60}[0.60]{\includegraphics{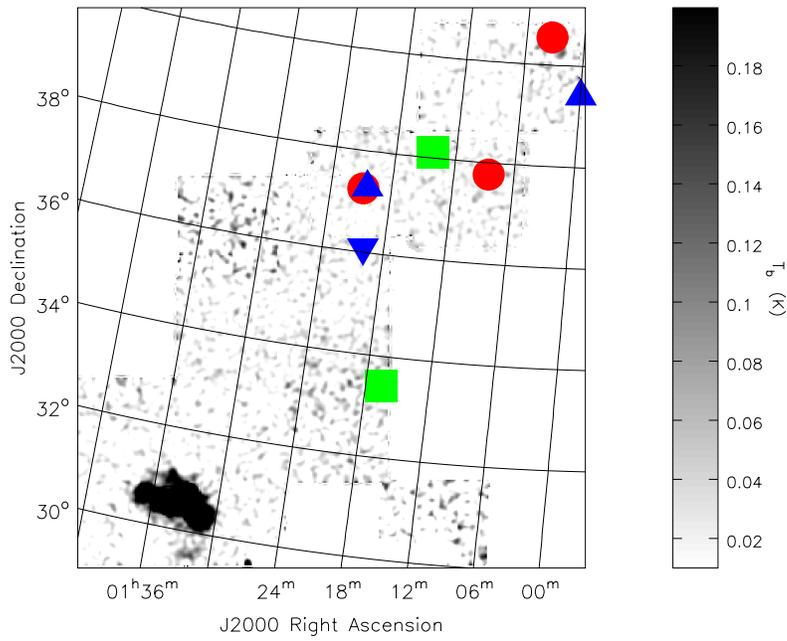}}
  \caption{
Area showing locations where \hi\ emission was detected in the survey (circles),  the location of the two deep pointings where \hi\ was detected  (triangles) and the one deep pointing without a detection (inverted triangle).  The two M31 dwarf galaxies covered by the survey are indicated by rectangles: And~XV to the north and And~II to the south.   M33 appears at the lower left.   We confirm the existence of \hi\ emission between M31 and M33 at several locations north of $\delta=36\arcdeg$, corresponding to a projected distance to M31 of $<120$ kpc.
}  
\label{fig:detection-fig} 
\end{figure}

\begin{figure}
 \scalebox{0.5}[0.5]{\includegraphics{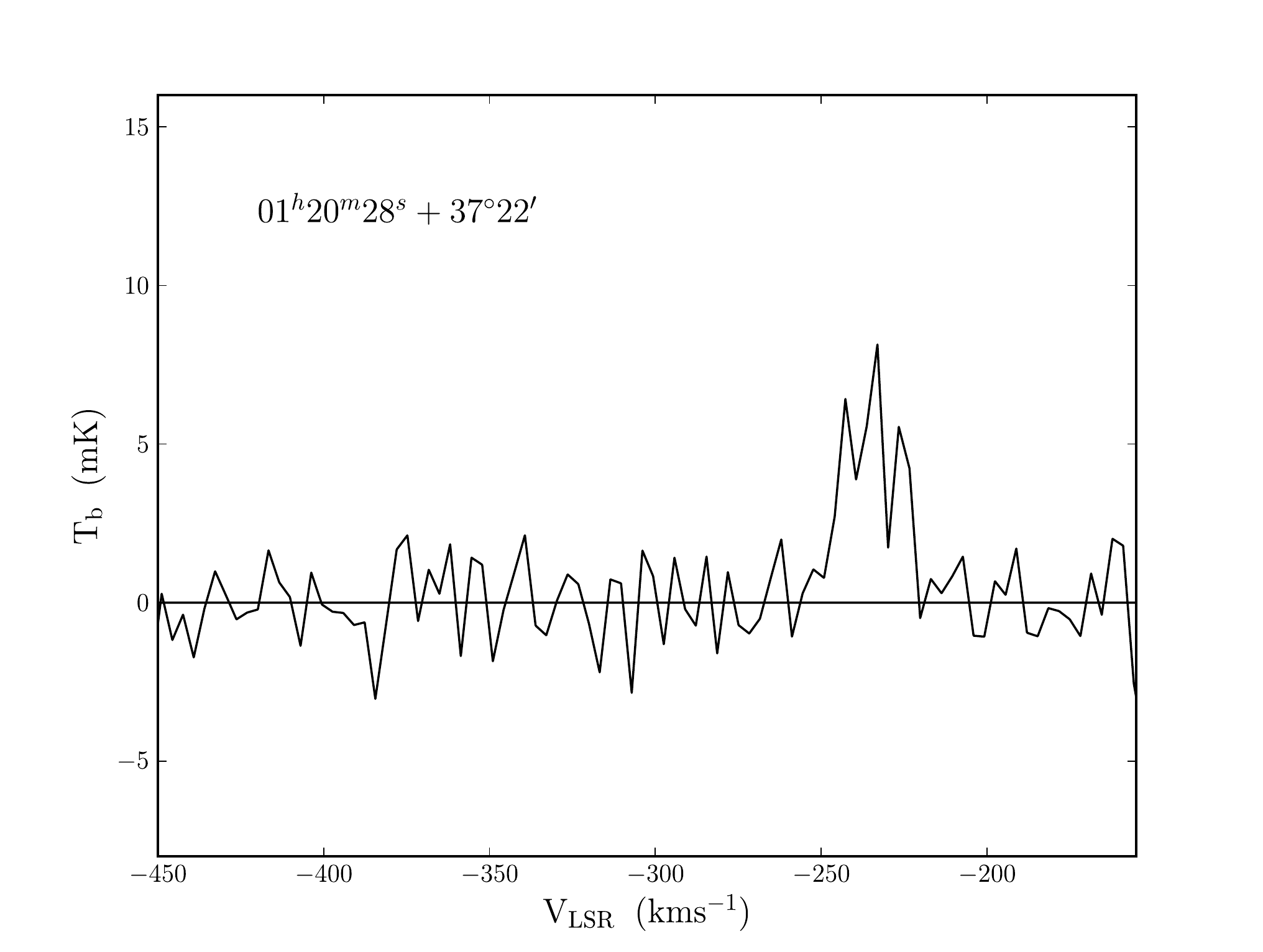}}
  \caption{A portion of the deep pointing spectrum toward $01^h20^m, +37\arcdeg22\arcmin$.  This resulted from 3.5 hours of in-band frequency switched observations acquired over several months.  The complete spectrum covers -600 to +400 km~s$^{-1}$. A  4th order polynomial instrumental baseline was removed.  The line at -235 km~s$^{-1}$ has a total N$_{\rm HI}$ of $2.5 \times 10^{17}$ cm$^{-2}$ (Table~\ref{tab:detections}).
}  
\label{fig:deepspectra_1}
\end{figure}

Fig.~\ref{fig:detection-fig} shows the survey area with various symbols indicating the location of  emission peaks detected in the map (circles), detections in the deep pointings (triangles), the deep pointing without a detected line (inverted triangle) and the two dwarf galaxies (rectangles).  
Properties of the lines are listed in Table~\ref{tab:detections} along with limits from the third deep pointing and fiducial information on M31 and M33.  
Line parameters were derived from a Gaussian fit, either to the follow-up spectra for lines detected in the map (indicated by the word ``Map'' in col.~8), or to the deep spectra.   
  The lines detected in the  deep pointings are shown 
 in Fig.~\ref{fig:deepspectra_1} and Fig.~\ref{fig:deepspectra_2}.    

\begin{figure}
 \scalebox{0.60}[0.60]{\includegraphics{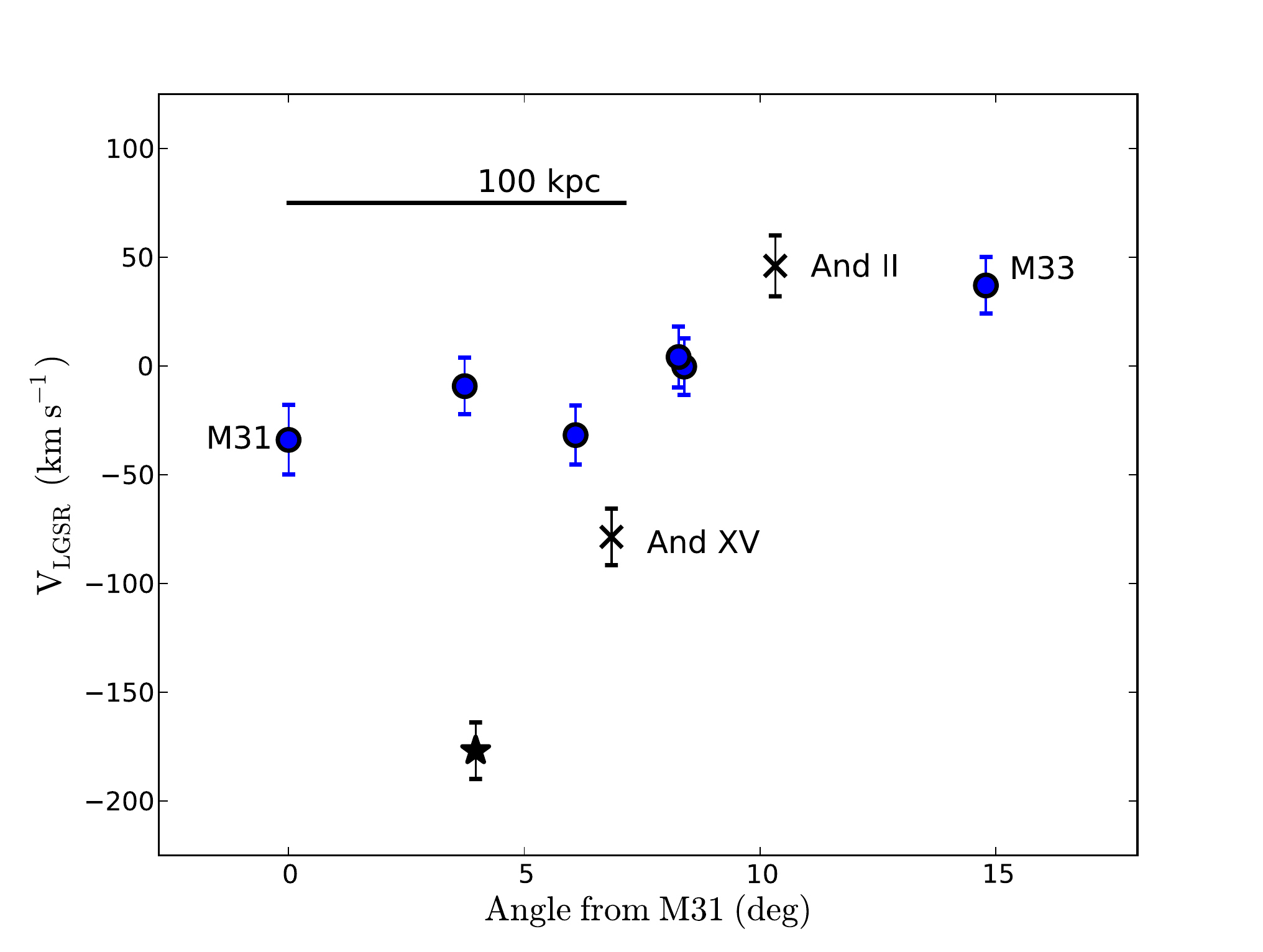}}
  \caption{Velocity with respect to the Local Group Standard of Rest (V$_{\rm LGSR}$) plotted against angle from M31 for the \hi\ lines detected between M31 and M33.  For reference, M31 and M33 are included.     Most detections have a velocity between that of  the two dominant galaxies.   The one discrepant point at V$_{\rm LGSR} = -187$ km~s$^{-1}$ (star)  is almost certainly unrelated emission from a cloud in the Magellanic Stream.
The dwarf galaxies And~II and And~XV are marked with crosses; they have no detectable \hi\ (Table \ref{tab:Satellites}).   The dominant source of velocity error is systematic and arises from uncertainty in the conversion of V$_{\rm LSR}$ to V$_{\rm LGSR}$; the experimental $1\sigma$ errors are $<2$ km~s$^{-1}$, about the size of the plotted symbols (see Table \ref{tab:detections}).  The bar shows the angle that corresponds to a projected distance of 100 kpc at the distance of M31, taken to be 0.8 Mpc.
}  
\label{fig:vlgsr}
\end{figure}

Velocities of the lines in the LGSR frame are plotted against angular
distance from M31 in Fig.~\ref{fig:vlgsr}. The errors are dominated by
the uncertainties in the conversion of V$_{\rm HEL}$ to V$_{\rm
LGSR}$.  Random errors in V$_{\rm LGSR}$ are typically the same order
as those in V$_{\rm LSR}$, a few km~s$^{-1}$.  With one exception, the
detections lie between the velocities and positions of M31 and M33, and
are thus likely related to these systems.  The exception is the line
at $01^h03^m21.9^s, +40\arcdeg33\arcmin$ which has a V$_{\rm LSR}$ and
a V$_{\rm LGSR}$ more than 100 km~s$^{-1}$ below that of the other
emission.  This gas is most likely related to an extension of the
Magellanic Stream, which has a similar velocity in this part of the
sky \citep[B\&T]{Nidever2010, Stanimirovic2008}, and not to M31 or
M33.  It is labeled as such in Table~\ref{tab:detections} and will not
be considered further here.

The brightest \hi\ lines detected in this survey have a column density about an order of magnitude larger than B\&T found at the same positions, and appear to be unresolved, or only slightly extended to the GBT beam.   Near $01^h20^m$ we have two detections, one from the map and one from a deep pointing taken fortuitously only $8\arcmin$ away.  If the map observation is centered on an unresolved \hi\ cloud the line brightness temperature at the offset position should be lower by a factor of 0.10.  The observed ratio of line intensities, T$_{\rm L}$,  is $0.071\pm0.013$.  Thus the observations are consistent, at the $3\sigma$ level, with the emission at $01^h20^m48.5^s, +37\arcdeg15\arcmin$ arising in an \hi\ cloud that is $<9\farcm1$ in angular size.  At a distance of 0.8 Mpc, the GBT beam has a linear size of  two kpc, implying \hi\ masses within the GBT beam of  $9.6 \times 10^4$ M$_{\odot}$  at $01^h20^m48.5^s,+37\arcdeg15\arcmin$, and $2.2\times 10^5$ M$_{\odot}$ at $01^h08^m32.5^s, +37\arcdeg46\arcmin$.

\section{The Bridge and the High Velocity Cloud Systems of M31 and M33}
\label{sec:HVCs}

Recent studies of M31 and M33 have refined our knowledge of their populations of HVCs, and it is interesting to compare the current detections with those objects.  The HVC system of M31 was discovered by \citet{Thilker2004} and investigated in depth by \citet{Westmeier2008}.  The HVCs are confined within a projected distance of 50 kpc despite sensitive searches of more distant areas;  two-thirds of M31's HVCs are within 30 kpc projected.   The situation is less clear for M33.  A list of possible HVCs has been assembled by \citet{Grossi2008} with several additions from \citet{Putman2009}, who note, however, that many from \citet{Grossi2008}  seem connected to that galaxy's gaseous halo.  

\begin{figure}
 \scalebox{0.60}[0.60]{\includegraphics{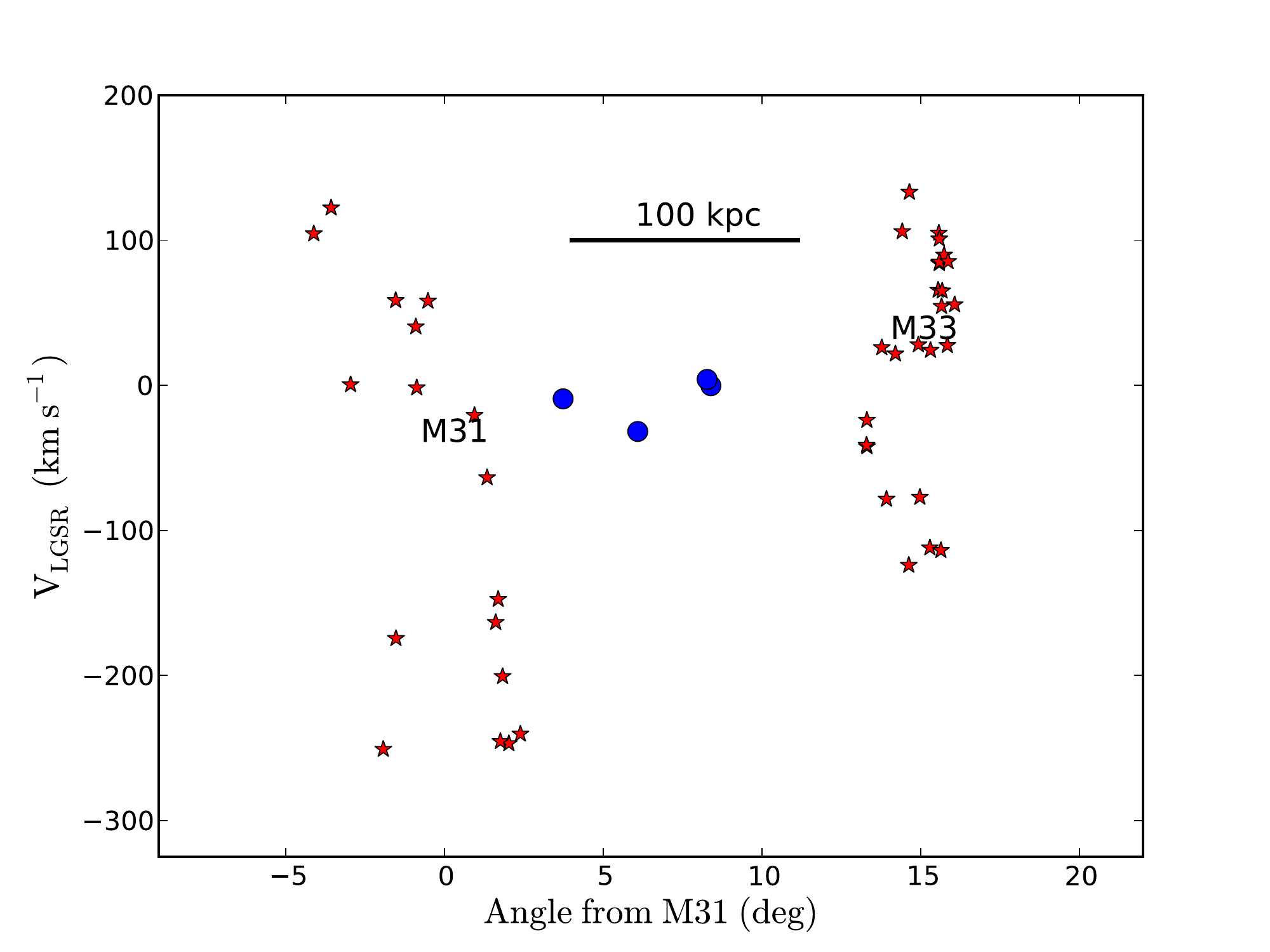}}
  \caption{Velocity with respect to the Local Group Standard of Rest (V$_{\rm LGSR}$) plotted against angle from M31 for the high velocity clouds around M31 and M33 (stars)  as well as the \hi\ lines detected from the bridge between between those galaxies (circles).   Measurement errors in velocity are comparable to the size of the symbols. The velocity and location of M31 and M33 are indicated, and the horizontal bar shows the angle subtended by 100 kpc at a distance of 0.8 Mpc.  The clouds detected in this survey appear related to the systemic velocity of the galaxies, and not to their populations of HVCs.  
}  
\label{fig:HVCs}
\end{figure}

The combined HVC populations of M31 and M33 are shown in
Figure~\ref{fig:HVCs} in angle from M31 vs. V$_{\rm LSGR}$, together
with the bridge emission features detected here (circles).  For
clarity, HVCs on the opposite side of M31 from M33 are given a negative
angular separation.  Whereas the HVCs around each galaxy show a wide
spread in velocity (those from M33 seem to trace a rotation curve) the
bridge clouds have a velocity near the systemic velocity of each
system.  The velocity dispersion of the M31 HVCs about  their mean 
V$_{\rm LGSR}$ is 130 km~s$^{-1}$, and for M33 is  
 either 76 or 50 km~s$^{-1}$, depending on whether
the entire sample or only those clouds from \citet{Putman2009} are used. 
For the bridge clouds the dispersion is  only 13 km~s$^{-1}$.  It is thus
probable that the bridge clouds arise from a very different source
than the HVCs. Their kinematics are consistent with B\&T's suggestion
that they form a partial link between the two galaxies.

\section{Concluding Discussion}
\label{sec:discussion}

Our observations have confirmed the existence of faint \hi\
concentrations between the galaxies M31 and M33.  They are found at
projected distances of 50-115 kpc from M31, at least half the distance to M33,
 have kinematics consistent
with the systemic velocity of M31 and M33, and appear to be distinct
from the HVC populations of the individual galaxies.  These results support the
basic discovery of \citet{BraunThilker2004}.  In two locations we
detect \hi\ an order of magnitude brighter than found by B\&T,
suggesting that the gas is highly clumped.  At one position the
emission is consistent with arising from a cloud with a size $<2$ kpc
and an \hi\ mass $\sim 10^5$ M$_{\odot}$.  This is similar to the size
and \hi\ content of the Milky Way dwarf galaxy Leo T
\citep{Irwin2007,Ryan-Weber2008}, although no stellar system has been
reported at the position of the \hi\ feature.

We do not detect \hi\ at one location in the southern-most extension of the B\&T bridge to a $5\sigma$ limit of N$_{\rm HI} \leq 1.5 \times 10^{17}$ cm$^{-2}$, inconsistent with the B\&T map at the $3\sigma$ level.  This implies that there is significant angular structure in the gas unresolved by the B\&T measurements, or that the bridge is confined to $\delta > 36\degr$, i.e., within 120 kpc of M31.  As all of our detections are near localized peaks in the B\&T map, we cannot confirm the existence of a very extended diffuse neutral \hi\ bridge  at levels $\sim 10^{17}$ cm$^{-2}$. 

No \hi\ was detected from the two known M31 dwarf galaxies covered by our survey, to limits of
 $\lesssim 10^4$ M$_{\odot}$.  This is not surprising, as both are dSph located less than 200 kpc from M31, a proximity that is correlated with an absence of significant \hi\ presumably because of  ram-pressure stripping in a hot halo \citep{Blitz2000,Grcevich2009}.  

\begin{figure}
 \scalebox{0.5}[0.5]{\includegraphics{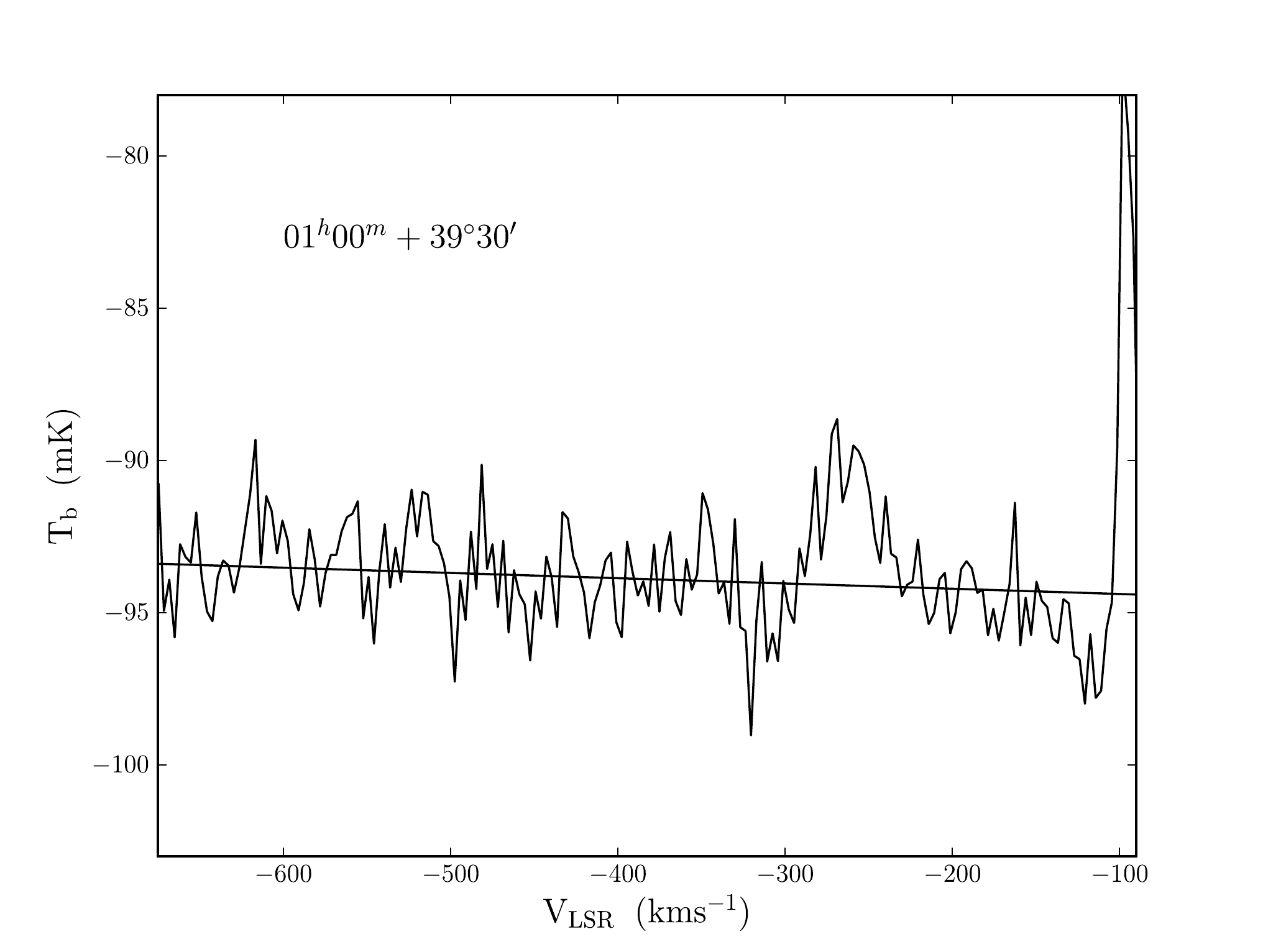}}
  \caption{A portion of the deep pointing spectrum toward $01^h00^m, +39\arcdeg30\arcmin$.  This resulted from about 6 hours of position-switched observations against a reference position at $3\degr$  higher right ascension.  The figure shows the calibrated data smoothed to 3.2 km~s$^{-1}$ velocity resolution without any correction for instrumental baseline.  The straight line was fit to the emission-free portions of the spectrum and indicates that the instrumental baseline is principly an offset with a slight slope.  The spectrum has an rms noise of 1.3 mK and the line at -262 km~s$^{-1}$ has a total N$_{\rm HI} = 2.9\pm0.2 \times 10^{17}$ cm$^{-2}$.  Complete properties are given in Table~\ref{tab:detections}.  At 
V$_{\rm LSR} > -150$ km~s$^{-1}$ the spectrum shows evidence of incomplete cancellation of extended emission that appears in both souce and reference positions.
}  
\label{fig:deepspectra_2}
\end{figure}

The \hi\ emission in the M31-M33 bridge is extremely faint and beyond the reach of most radio telescopes because of limitations on sensitivity and the quality of  instrumental baselines.  The  GBT spectrum in Fig.~\ref{fig:deepspectra_2} is among the highest-quality 21cm \hi\ emission spectra ever 
obtained at this low noise level.  Further 21cm observations with the GBT are planned to study the structure and kinematics of the M31-M33 bridge.  It would also be extremely interesting to measure the bridge in UV
absorption lines against distant AGN to gain information on its metallicity and ionization stage,  and the amount of ionized gas that is likely associated with the structure.

\acknowledgements

We thank the anonymous referee for useful suggestions.

{\it Facility:} \facility{GBT}

\begin{deluxetable}{lclcccll}
\tabletypesize{\small}
\rotate
  \tablecaption{Summary of Measurements of the M31-M33 Bridge 
 \label{tab:detections}}
\tablehead{
    \colhead{J2000} & \colhead{T$_{\rm L}$} &  \colhead{FWHM} &  \colhead{V$_{\rm LSR}$} & \colhead{$\sigma_{\rm b}$\tablenotemark{1}} & \colhead{N$_{\rm HI}$}
& \colhead{V$_{\rm LGSR}$} & \colhead{Notes} \\
\colhead{(hh:mm:ss.s dd:mm)} & \colhead{(mK)} & \colhead{(km s$^{-1}$)}  & 
\colhead{(km s$^{-1}$)} & \colhead{(mK)} & \colhead{(10$^{17}$ cm$^{-2}$)} & 
\colhead{(km s$^{-1}$)} & \\
\colhead{(1)} & \colhead{(2)}  & \colhead{(3)}  & \colhead{(4)}  & \colhead{(5)} & \colhead{(6)} & 
\colhead{(7)} & \colhead{(8)} \\ }
  \startdata
01:00:00.0 +39:30 & $4.4 (0.4)$ & $34.4 (4.0)$ & $-262.4 (1.7)$ & 1.3 & $2.9 (0.2)$  & $-9$ (13)  
& Deep \\
01:03:21.9 +40:33 & 86 (4) & 24.1 (1.2) & $-430$ (0.5) & 8.7 & 40 (1) & $-177$ (13) & 
Map\tablenotemark{2}  \\
01:08:32.5 +37:46 & 106 (3) & 38.0 (1.2) & $-278$ (0.5) & 8.6 & 78 (1) & $-32$ (13.5) & Map \\
01:20:00.0 +36:00 & $\leq 9$ &              &                       & 1.8  & $\leq 1.5$ &                 & Deep \\
01:20:28.3 +37:22 & 6.1 (0.9) & 20.8 (3.6) & $-235.1$ (1.5) & 1.3 & 2.5 (0.2) & $+4$  (14) & 
Deep \\
01:20:48.5 +37:15 & 75 (4) & 23.3 (1.3) & $-239$ (0.6) &  7.2 & 34 (1) & $-0.3$  (13) & 
Map \\
00:42:44.3 +41:16 &         &        &  $-296$ (4)  &  &     & $-34$ (16)    & M31 \\
01:33:50.9 +30:39 &         &        &  $-180$ (3)  &  &     & $+37$ (13)    & M33 \\
 \enddata 
\tablecomments{Uncertainties are $1\sigma$, limits are $5\sigma$. Values for M31 and M33 were taken from NED: http://ned.ipac.caltech.edu.  Conversions from V$_{\rm LSR}$ to V$_{\rm LGSR}$ 
were made using the apex velocity and coordinates given by \citet{Karachentsev1996}.}
\tablenotetext{1}{RMS brightness temperature noise in a 3.2 km s$^{-1}$ channel.}
\tablenotetext{2}{Probably part of the Magellanic Stream.}
\end{deluxetable}




\end{document}